\documentclass[11pt,twoside]{article}
\usepackage{./asp2014}

\aspSuppressVolSlug
\resetcounters

\begin{document}

\title{Stellar atmosphere interpolator for empirical and synthetic spectra}
\author{Nikolay~Podorvanyuk,$^1$Igor~Chilingarian$^{2,1}$ and Ivan~Katkov$^1$
\affil{$^1$Lomonosov Moscow State University, Sternberg Astronomical Institute,
13 Universitetsky prospect, Moscow, Russia, 119991; \email{nicola@sai.msu.ru}}
\affil{$^2$Smithsonian Astrophysical Observatory, 50 Garden St. MS98, Cambridge, MA 02138, USA}}

\begin{abstract}
We present a new stellar atmosphere interpolator which we will use to
compute stellar population models based on empirical and/or synthetic
spectra. We combined observed and synthetic stellar spectra in order to
achieve more or less uniform coverage of the ($T_{eff}, \log g, [Fe/H]$) 
parameter space. We validated our semi-empirical stellar
population models by fitting spectra of early-type galaxies from the SDSS
survey.
\end{abstract}

\section{Motivation}

Empirical stellar spectra (e.g. ELODIE \citep{PS04} and MILES
\citep{SanchezBlazquez+06} stellar libraries) are broadly used in stellar
population synthesis codes, however they do not cover the
temperature--gravity--metallicity ($T_{eff}, \log g, [Fe/H]$) parameter
space uniformly that requires extrapolation in the parameter space and leads
to artefacts in stellar population models computed with empirical stellar
libraries.  Those gaps can be filled using synthetic stellar atmospheres
(e.g., PHOENIX \citealp{Husser+13} and
BT-Settl\footnote{\url{https://phoenix.ens-lyon.fr/Grids/BT-Settl/}}).  However, using fully synthetic grids
for stellar population synthesis is not recommended because they suffer from
incomplete line lists and do not reproduce observed integrated light spectra
well enough.

\articlefigure{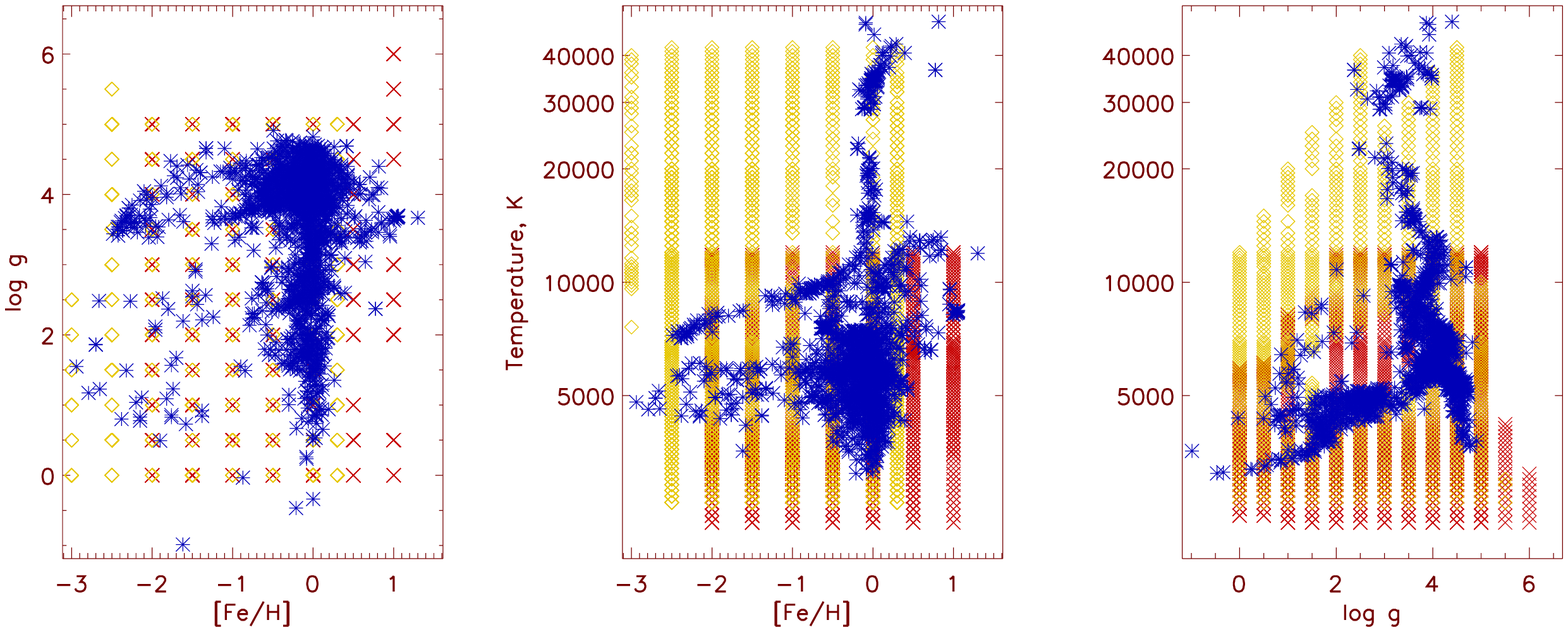}{ex_fig1}{The parameter space for empirical
(ELODIE - blue points) and synthetic (PHOENIX - red and BT-Settl - yellow)
stellar libraries: (a) $T_{eff} - [Fe/H]$, (b) $[Fe/H] - \log g$, (c)
$\log g - T_{eff}$.  These three libraries cover the full range of parameters
required by stellar population fitting codes.}

\section{Implementation}

We propose a new interpolation procedure that is based on a non-parametric
smoothing of fluxes from stellar spectra in the ($T_{eff}, \log g, [Fe/H]$)
parameter space. We apply our interpolation procedure at every spectral
pixel. Before we run our procedure, we normalize all spectra to the unity at
the wavelength 5500\AA.

First, we combine libraries of empirical and synthetic stellar spectra. This
is done by analyzing the distribution of observed stars in a 3-dimensional
parameter space and creating a concave hull around them.  A concave hull is
similar to a convex hull, a uniquely defined convex polyhedron that can be
constructed for any distribution of points in a multidimensional space and
will contain all the points while all its vertices will also be the points
from the same set.  A concave hull is a non-convex polyhedron that possesses
the same properties regarding the inclusion but excludes ``cavities'' in the
space that do not contain any points from the original distribution.  The
maximal characteristic size of those cavities is defined by an additional
parameter.

For observed Milky Way stars in the stellar atmosphere parameter space,
there is a pronounced cavity at high temperatures, low metallicities and low
surface gravities (see Fig.~\ref{ex_fig1}). We take synthetic stars outside
the concave hull constructed around empirical stellar spectra and add them
to our dataset in order to avoid extrapolation when computing interpolated
stellar spectra. We do this, because we do not want the situation when 
synthetic atmospheres dominate in the region of the parameter space where
observed spectra are available.

Then, we use smoothing splines (basic splines or $b$-spline) on $T_{eff}$
and a low-order two-dimensional polynomial fitting surface on $\log g$ and
$[Fe/H]$ to approximate the distribution of fluxes from corresponding
stellar spectra normalized to the unity at the wavelength 5500\AA.  The use
of the smoothing function is essential because: (i) there are uncertainties
in the stellar atmosphere parameter determinations for real stars; (ii)
there are physical star-to-star variations in the chemical composition that
cause spectra of several stars having very close values of atmosphere
parameters to differ from each other.

The obtained parametrization is then evaluated at every point of the
parameter space of the output [regular] stellar atmosphere grid.  Since our
goal is to compute stellar population models with the {\sc pegase.hr}
evolutionary synthesis code, we use the same grid of atmosphere parameters
as the one taken by the code.  We obtain the flux re-normalization values
for our spectra by measuring them at $\lambda=5000$~\AA\ in one of the
synthetic stellar atmosphere grids provided within the {\sc pegase.hr}
package.

We also evaluate the parametrization at every point of the input stellar
atmosphere grid in order to check the interpolation quality and detect
possible systematics. In Fig.~\ref{ex_fig2} we show the standard deviation of
the residuals between the input stellar spectra and interpolated spectra at
the same positions of the parameter space. 

The use of$b$-splines helps us to resolve a long standing issue in the
stellar population modelling regarding a mathematically correct way of
stellar atmosphere interpolation that does not cause discontinuities in
resulting stellar population models, which hampers stellar population
analysis in real galaxies and star clusters.

\articlefigure{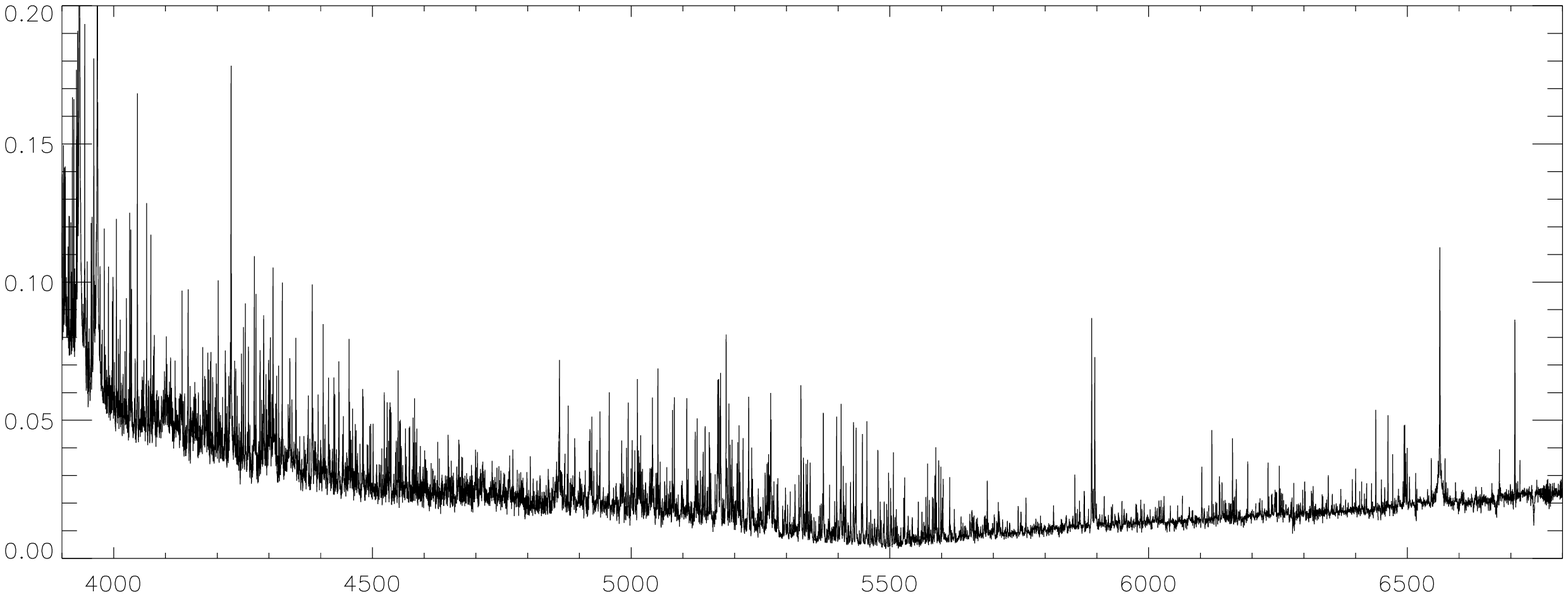}{ex_fig2}{Standard deviation of the difference
between interpolated and original spectra at every point in wavelength
range.}

\section{Results}

We computed a grid of stellar population models by feeding our
interpolated semi-empirical stellar atmosphere grid to the {\sc pegase.hr} population
synthesis code \citep{LeBorgne+04}. We tested the new grid of models by fitting 
spectra of galaxies from the SDSS survey \citep{SDSS_DR7}. 

In Fig.~\ref{ex_fig3} we show an example of the spectral fitting for the
galaxy 51913-0394-625.  We fitted its spectrum with a new technique for the
determination of the initial mass function in unresolved stellar populations
\citep{NP2013}.  We see that the fitting residuals are similar compared to
the results obtained by fitting the same galaxy using the original {\sc
nbursts} code \citep{CPSK07,CPSA07} with ELODIE-based stellar population
models (Fig.~\ref{ex_fig4}).

\articlefigure{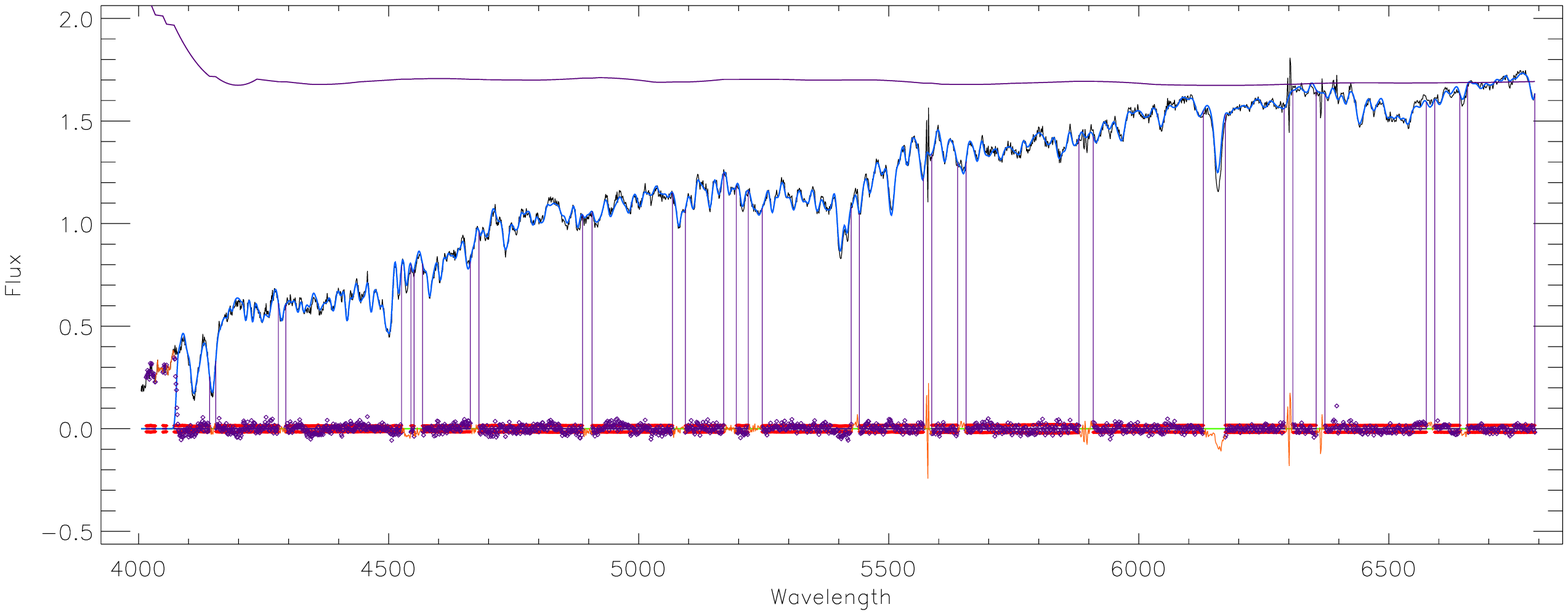}{ex_fig3}{Galaxy 51913-0394-625 from SDSS survey,
fitted by semi-empirical stellar population models using ELODIE and PHOENIX
stellar libraries.  Upper line shows the continuum shape.  Several regions
(e.g., night-sky emission lines and Na{\sc i} D) were excluded from the
fitting.}

\articlefigure{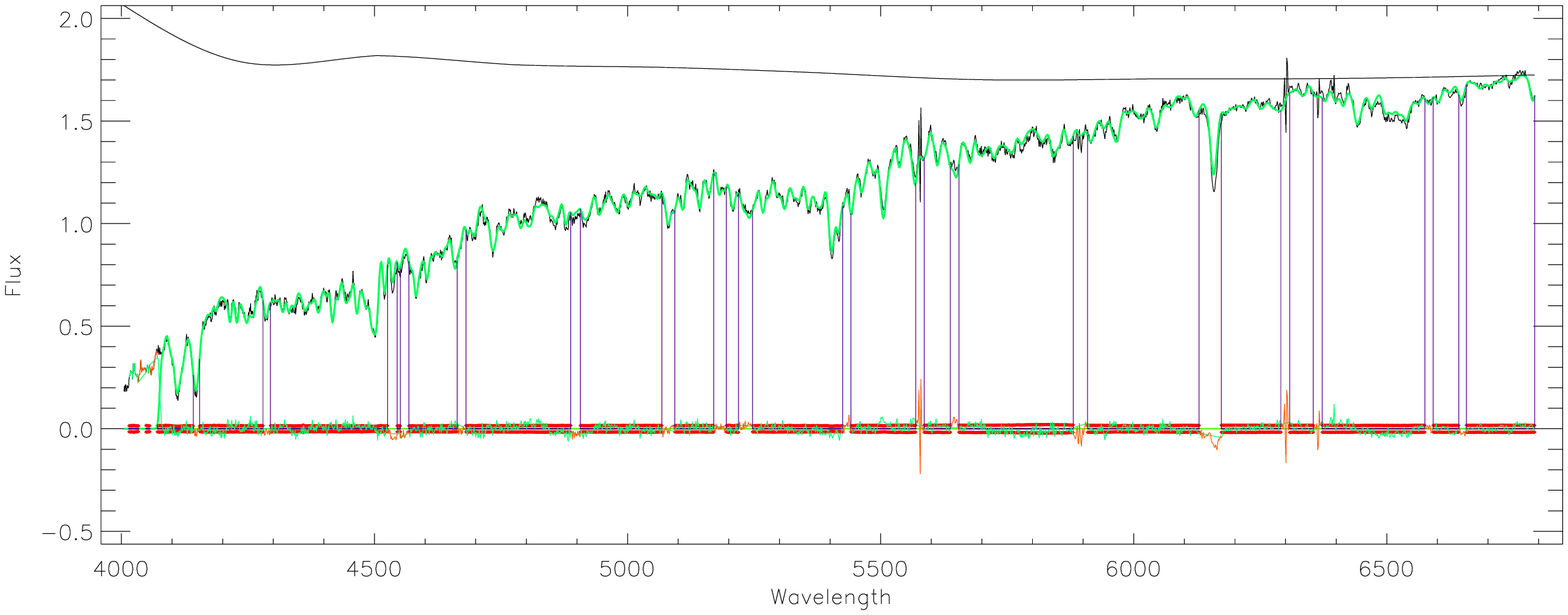}{ex_fig4}{Galaxy 51913-0394-625 from SDSS survey,
fitted using ELODIE based {\sc pegase.hr} stellar population models.  Upper
line shows the continuum shape.}

We analysed spectral data for 3000 giant elliptical galaxies from the SDSS
survey and other galaxies using our full spectrum fitting technique.  The
detailed comparison of our results (ages, metallicities) for different grids
of stellar population models will be provided in a separate paper (Katkov et
al. in prep.)
	
\acknowledgements This work is supported by a grant of the President of
the Russian Federation MD-7355.2015.2, and Russian Foundation for Basic
Research grants 15-52-15050 and 15-32-21062.

\bibliography{P090}  % For BibTex

\end{document}